\documentclass[prb,twocolumn,showpacs,floatfix]{revtex4-1}
\usepackage[dvips]{graphicx}
\usepackage{epsfig}
\usepackage{color}
\usepackage[normalem]{ulem}
\usepackage{amsmath,amsfonts,amssymb,bm}
\setcitestyle{numbers,square}
\usepackage{verbatim}
\usepackage[capitalize]{cleveref}
\usepackage{color,soul}
\begin{document}
\title{Many-body localization in spin chains with the long-range transverse interactions: scaling of critical disorder with the system size}
\author{Andrii O. Maksymov,  Alexander L. Burin}
\affiliation{Tulane University, New
Orleans, LA 70118, USA}

\date{\today}
\begin{abstract}
We investigate many-body localization in the chain of interacting spins with a transverse power-law interaction, $J_{0}/r^{\alpha}$, and random on-site potentials, $\phi_i \in \left(-W/2,W/2\right)$, in the long-range limit, $\alpha< 3/2$, which has been recently examined experimentally on trapped ions. The many-body localization threshold is characterized by the critical disordering, $W_c$, which separates localized ($W > W_c$) and chaotic ($W < W_c$) phases. Using the analysis of the instability of localized states with respect to resonant interactions complemented by numerical finite size scaling, we show that the critical disordering scales with the number of spins, $N$, as $W_c \approx [1.37 J_{0}/(4/3 - \alpha)]N^{4/3 - \alpha} \ln N$ for $0 < \alpha \leq 1$, and as $W_c \approx [J_{0}/(1-2\alpha/3)]N^{1-2\alpha/3} \ln^{2/3} N$ for $1 < \alpha < 3/2$ while the transition width scales as $\sigma_{W} \propto W_{c}/N$. We use this result to predict the spin long-term evolution for a very large number of spins ($N = 50$), inaccessible for exact diagonalization, and to suggest the rescaling of hopping interaction with the system size to attain the localization transition at finite disordering in the thermodynamic limit of infinite number of spins. 
\end{abstract}
\maketitle

% Changes references Monro16 and Smith2016  were to the same work so I put them together. Replaced reference 9 with that to Monroe where the modification of exponent has also been discussed. Added reference to Mossner17LR

\section{Introduction}

Many-body localization (MBL) transition separates two distinguishable kinetic behaviors: the delocalized, chaotic system acts as a thermal bath for each small part of it  \cite{Huse15Thermalization,Cohen13}, while in the localized system its different parts are  approximately independent. In the chaotic phase, energy levels obey the Wigner-Dyson statistics, while in the localized phase, they obey the Poisson statistics \cite{OganesyanHuse07}.  The localized phase can be characterized by related local integrals of motion  \cite{Serbyn13LocalIntMot,Huse14IntMot} (see also Refs. \cite{Huse15Thermalization,Serbyn19Review} for review of a more recent progress in this area). The experimental investigations of many-body localization \cite{Monro16,LukinDiamond16,Debnath2016,Atoms51Bernien2017,Atoms51Bernien2017,Britton2012,Edwards2010,Saffman2010} are carried out in systems of interacting spins coupled by the long-range interaction, which decreases with distance according to the power law $U(r) \propto r^{-\alpha}$. The interaction exponent, $\alpha$, can be modified experimentally \cite{Yao14MBLLongRange,Monro16}, which helps to understand the effect of different power-law interactions on localization. Such systems are of interest particularly due to their relevance in quantum computing \cite{Debnath2016,Britton2012},  while the ubiquitous power-law interactions are associated with the presence of dipole, magnetic or elastic moments \cite{YuLeggett88,ab98book,ab06preprint,Douglas2015,Yan2013,Otten2016}. Theoretical studies of these systems include the investigation of entanglement entropy \cite{Mossner17LR}, superdiffusive transport \cite{BarLev15}, ultrafast propagation of information \cite{Tran2019} and search for many-body localization in various settings \cite{Nandkishore17lr,Li16,GargSK18,Logan19,Garg19,Zakrzewski19,Maze2011}.

It has been recently shown experimentally \cite{Monro16,Atoms53Zhang2017} that the system of $N$ spins with long-range power-law interactions can be modeled on trapped ions simulating Hamiltonians  of the form
\begin{eqnarray}
\hat{H}=\sum_{i<j}J_{ij}\sigma_{i}^{x}\sigma_{j}^{x}+\frac{1}{2}\sum_{i}\left(B+\phi_i\right)\sigma_{i}^{z},
\label{eq:H}
\end{eqnarray}
where $J_{ij}=J_0|j-i|^{-\alpha}$ is the long-range interaction with a tunable exponent $\alpha$. Random fields, $\phi_i$, are uniformly distributed in the range $(-\frac{W}{2},\frac{W}{2})$, with an additional transverse field, $B$, added to make the system delocalized in the absence of disordering. The system is expected to be localized at sufficiently strong disordering, $W$, where interaction can be neglected, and the eigenstates are defined by spin projection operators $S^{z}$, which serve as local integrals of motion. The localization threshold is determined by the critical disordering $W_{c}$ such that the states are localized for $W>W_{c}$ and delocalized otherwise. 

%Following Ref. \cite{Monro16} we investigate the system described by Eq. \eqref{eq:H} with the main consideration given to the case $B=4J_{0}$ where the localized regime is realized at $W \gg 1$ only. The case $B=0$ is considered for comparison for few power law exponents $\alpha$. In the latter case the system has a localization transition also at very small disordering which can be described following Ref. \cite{Burin2017} so it is not the target of the present work (see Sec. \ref{sec:Num} for details).  

Many-body localization breakdown due to the instability of localized states with respect to resonant long-range interactions has been considered in the earlier work \cite{ab89TLSJETP,ab98prl,ab98book,ab06preprint,ab15MBL}, in models with both off-diagonal (transverse, hopping $\sigma_i^x\sigma_j^x$) and diagonal (longitudinal, $\sigma_i^z\sigma_j^z$) power-law interactions with exponents $\alpha$ and $\beta \leq \alpha$, respectively. Similar premises were used for electronic systems in Ref. \cite{ab16GutmanMirlin,abGorniyMirlinDot}. According to those considerations, delocalization takes place in the presence of around one resonance if the diagonal interaction of resonant transitions exceeds their amplitude. This consideration led to the dimensional constraint $\beta<2d$, for which delocalization always takes place in the thermodynamic limit of infinite $N$. Recent papers \cite{Logan19,Garg19,Zakrzewski19} challenging this constraint possess isotropic interaction. According to Ref. \cite{Yao14MBLLongRange} such interaction leads to a much weaker dimensional constraint ($\beta+2<2d$), which turns out to be  consistent with the numerical results of Refs. \cite{Logan19,Garg19}. %This dimensional constraint has been generalized to systems with isotropic diagonal interaction and fixed total spin projection (or fixed number of particles) \cite{Yao14MBLLongRange}  by modifying the dimensional constraint to $d>\alpha(\beta+2)/(\alpha+\beta+4)$, while in systems without such conservation laws like in Eq. (\ref{eq:H}), interactions generally can be  treated as anisotropic.

The argument of resonant interactions \cite{ab06preprint,Yao14MBLLongRange,ab15MBL} is not applicable to the system described by Eq. (\ref{eq:H}) since it lacks the diagonal interaction. The investigation of the X-Y model lacking the longitudinal interaction \cite{ab15MBLXY} led to a weaker dimensional constraint, $\alpha< 3d/2$. The power-law scaling of the critical disordering, $W_{c}$, with the number of spins, $N$, was predicted there for $d<\alpha<3d/2$.  

The main target of the present work is to determine the critical disordering in the regime of violated dimensional constraint $\alpha < 3d/2$ in the model described by Eq. (\ref{eq:H}). Using the analysis of the instability of localized states with respect to resonant interactions complemented by numerical finite size scaling, we express  the critical disordering as an algebraic function of the number of spins, $N$. This expression can be used to characterize arbitrarily large spin systems, including those not accessible for numerical simulations but only for experimental measurements ($N \approx 50$ \cite{Atoms53Zhang2017,Atoms51Bernien2017}). We demonstrate that the most efficient delocalization is associated with resonant spin quartets. The analysis of the localization breakdown by quartets leaves dimensional constraint of Ref. \cite{ab15MBLXY} ($\alpha<3d/2$) unchanged, but leads to a faster increase of the critical disordering with the system size.  It should be easier to observe the latter scaling in practice, compared to the very slow increase of the critical disordering predicted in Ref. \cite{ab15MBLXY}, as noticed in Ref. \cite{Nandkishore2018}. The localization threshold is also determined for the power-law interaction with small exponents $0<\alpha <1$, which can also be realized in cold ions \cite{Monro16,Atoms51Bernien2017,Atoms53Zhang2017}.

The recent work \cite{HotSpotLuitz2017,Huveneers17BrekDwnLoc} suggests that the power-law interaction always breaks down MBL at sufficiently large system sizes, because of chaotic spots; however, the critical disordering is expected to increase logarithmically with the system size \cite{Mirlin16Fractal,HotSpotLuitz2017,Huse2019Arxiv}. This dependence is weaker than the power-law scaling; therefore, the predictions of the present work for the localization threshold in the case of $\alpha< 3d/2$ should remain valid.

 %{\bf Similar models has been studied in other work \cite{Mossner17LR,Garg19,Logan19}. Particularly, the analytical and numerical considerations in Refs. \cite{Garg19,Logan19} led to a different dimensional constraints $\alpha<2$ and $\alpha<1/2$.}

The paper is organized as following. In Sec.  \ref{sec:Scal}, we derive the scaling of the critical disordering with the system size up to the accuracy of a power law and logarithmic factors, based on the consideration of the delocalization induced by resonant spin quartets. The obtained dependence is then used to suggest rescaling of the interaction constant leading to the finite localization threshold within the thermodynamic limit of infinite number of spins. We show that this threshold is stable with respect to the higher order resonance (sextets, etc.) and chaotic spots \cite{Huveneers17BrekDwnLoc}.

In Sec. \ref{sec:Num}  we compare numerical results for Hamming distance and level statistics obtained by means of exact diagonalization with some experimental data \cite{Monro16} and our analytical predictions, using the latter comparison to determine numerical factors in the definition of the critical disordering. We also obtain a universal expression for the transition width, $\sigma_W$, justifying it both numerically and analytically.

We summarize our results in Sec. \ref{sec:Concl} and make a prediction for the Hamming distance in a system of N=50 spins with the power-law interaction exponent $\alpha=0.5$. This prediction can be directly compared to the Hamming distance measurements like in Ref. \cite{Monro16}. 

Appendices A and B include detailed derivations of the localization threshold due to interacting spin pairs and spin quartet transition amplitudes.

%Experimental investigations of MBL in spin systems with long-range interactions \cite{Monro16} has stimulated numerous research efforts targeting many-body localization and dynamics in systems with long-range interactions. The latter includes the investigation of entanglement entropy showing anomalous time dependence \cite{Mossner17LR}, superdiffusive transport  \cite{BarLev15} and many-body localization in various settings \cite{Li16,GargSK18,Logan19,Garg19,Zakrzewski19}.  Despite some works challenging the dimensional constraint on the basis of  the reported numerical and analytical results \cite{Logan19,Garg19,Zakrzewski19}, we think that for the studied systems with isotropic interactions and fixed total spin projection (or fixed number of particles) a modified dimensional constraint \cite{Yao14MBLLongRange} should be used instead as briefly discussed in Appendix C.

%%%%%%%%%%%
% Disagree that the discussion of de Roeck should go to the end because this is the important result that put us under the question. 

\section{Delocalization due to interacting quartets.}
\label{sec:Scal}

In the recent paper \cite{ab15MBLXY}, the many-body localization transition has been considered within the X-Y model with $1/r^\alpha$ interaction. The resonances in the pairs of interacting spins leading to the delocalization has been examined as a potential source of delocalization, while the longitudinal interaction between them has been generated in the third order of perturbation theory. This consideration has lead to the dimensional constraint for localization requiring $\alpha>3d/2$, which converts to $\alpha>3/2$ in the case of interest for $d=1$. Here and in the rest of the paper we consider one-dimensional systems except for the few cases targeting the generalization to higher dimensions.  

The scaling of localization threshold has been obtained in the form $W_{c}\propto N^{\frac{\alpha(3-2\alpha)}{2(\alpha+1)}}$. Here, we demonstrate that the interacting resonant transitions of spin quartets  lead to more efficient delocalization and lower critical disordering, $W_{c}$. The consideration is also generalized to the practically significant case of the smaller power law interaction exponents $0<\alpha \leq 1$, where delocalization is also determined by interacting resonances in spin quartets, as can be seen by comparing the criteria for spin quartets (derived below) with the criteria for spin pairs (derived in Appendix A). 

\subsection{General definition of the localization threshold.}

We use the definition of delocalization transition of Ref. \cite{Burin2017} for the system of $N$ spins, coupled by the long-range interaction, in the way that the interaction at the maximum distance is the most significant, and the most relevant multi-spin resonances are associated with spins separated by the system size. We will see below that this is the case for interacting spin quartets in the problem given by Eq. \eqref{eq:H}, in the regime of interest of $\alpha < 3/2$ ($3d/2$ in a d-dimensional system). 

A collective transition of $k$ spins ($k=2$ for pairs and $k=4$ for quartets of spins) can be characterized by the transition amplitude $V_{*}$. The probability of resonance for a single $k$-spin transition can be expressed as $V_{*}/W$. There are approximately $N_{*} \sim N^{k}/k!$ ways to create $k$-spin excitations, so the total number of resonances can be estimated as $N_{*}V_{*}/W$, averaged over possible realization of the $k$-spin transition. According to Ref. \cite{Burin2017}, it is sufficient to have few resonances per system, if the diagonal interaction between them, $U_{*}$, exceeds the typical resonant coupling strength, $V_{*}$. Then the delocalization transition can be described using the similarity of the present problem with the localization problem on the Bethe lattice, with resonant coupling $V_{*}$, disordering $W$, and coordination number $N_{*}$ \cite{AbouChacra73}. 
The transition is detemined as \cite{Burin2017}
\begin{eqnarray}
N_*\frac{V_*}{W}\ln \frac{U_*}{V_*}  \approx 1. 
\label{eq:resthr}
\end{eqnarray}
Since amplitudes $V_{*}$ for different spin transitions can fluctuate, one can define the typical amplitude $V_{*}$ as the average absolute value of contributing amplitudes similarly to Refs. \cite{AbouChacra73,Burin2017}. 

This criterion is used below to describe the delocalization due to quartet spin transitions. Considering all quartets of spins, characterized by transition amplitudes $V_{klmn}$, defined below in Sec. \ref{quartets}, one can rewrite the criterion of Eq. \eqref{eq:resthr} in the form
\begin{align}
        \frac{\eta}{W}\sideset{}{'}\sum_{klmn}\left<\left|V_{klmn}\right|\right>\ln \left( \frac{U_*}{\left<\left|V_{klmn}\right|\right>} \right)\approx 1,
    \label{eq:Vklmn}
    \end{align}
where the prime means that the sum is only over quartets with $V_{klmn}<U_*$, and the diagonal interaction of resonances, $U_*$, estimated in Sec. \ref{diagonal}. We do not target analytically the unknown numerical constant  $\eta \sim 1$ in Eq. \eqref{eq:resthr}, because there are correlations in various contributions to the transition amplitude, $V_{klmn}$, which are too difficult for accurate analytical calculations; instead, we determine that constant using the numerical study of the same problem  in Sec. \ref{sec:Num}. 

\subsection{Definition of quartets and their coupling amplitudes} \label{quartets}
Resonances created by interacting clusters of four spins should be considered after the resonances of spin pairs, since only transitions of even numbers of spins are permitted in the system described by Eq. \eqref{eq:H}. Flipping a spin quartet from an all up to an all down state can be done by flipping each pair of spins independently in a second order process. That process, however, has zero amplitude in the resonant regime $\phi_{1}+\phi_{2}+\phi_{3}+\phi_{4}=0$ (see Appendix B), due to the destructive interference of processes like
\begin{align}
       \left|\uparrow\uparrow\uparrow\uparrow\right> \rightarrow \left|\downarrow\downarrow\uparrow\uparrow\right> \rightarrow \left|\downarrow\downarrow\downarrow\downarrow\right>.
\end{align}
For the sake of simplicity, we consider the case of zero external constant field, $B=0$, in the Hamiltonian \eqref{eq:H}. Since the critical disorder approaches infinity with increasing the system size, the results in that limit should not be sensitive to the finite field $B$ that is consistent even with our finite size numerical studies in Sec. \ref{sec:Num}. 

The perturbation theory can be used to estimate spin-quartet transition amplitudes, $V_{klmn}$, if there are no resonant spin pairs within the quartet. Resonances can be excluded by setting a constraint, $|\phi_{i}-\phi_{j}|>J_{ij}$, that is justified with logarithmic accuracy, similarly to Refs. \cite{AbouChacra73,Anderson58,Burin2017}.
The transition amplitude, $V_{klmn}$, comprises four distinguishable contributions,
\begin{align}
   V_{klmn}= A_{k,lmn}+A_{l,kmn}+A_{m,kln}+A_{n,klm},
\end{align}
%led by each spin flipping three times with each other spin as shown on the diagram
%A non-zero amplitude appears in the third order when rotating every spin together with one selected spin as
where each contribution corresponds to a process led by one of four spins, flipping three times with other spins, as illustrated below for the representative process led by the first spin:
    \begin{align}
       \left|\uparrow\uparrow\uparrow\uparrow\right> \rightarrow \left|\downarrow\downarrow\uparrow\uparrow\right> \rightarrow \left|\uparrow\downarrow\downarrow\uparrow\right> \rightarrow \left|\downarrow\downarrow\downarrow\downarrow\right>.
       \label{eq:proc}
    \end{align}
Each individual contribution can be expressed as (see Appendix B)
%The amplitude of the four-spin transition can be expressed using the perturbation theory; enumerating spins with the letters $k, l, m, n$, one can express this amplitude as a sum over six processes, like Eq. \eqref{eq:proc} (see appendix B), which contribute to the flip of the $klmn$ spin quartet, with the same coefficient through one selected spin $k$ as
    \begin{multline}
        A_{k,lmn}=\sum_{\{lmn\}} \frac{J_{kl}J_{km}J_{kn}}{\left(\phi_k+\phi_l\right)\left(\phi_l+\phi_m\right)}= \\ =\frac{4\phi_{k}J_{kl}J_{km}J_{kn}}{(\phi_{l}+\phi_{m})(\phi_{l}+\phi_{n})(\phi_{m}+\phi_{n})},
    \label{eq:Aklmn}
    \end{multline}
where the sum is taken over all six permutations of the indexes $\{lmn\}$.

%We can see that the result is not zero, and  three other channels would give contributions proportional to the product of different amplitudes, so the overall amplitude is generally not zero, except for the case when all amplitudes are equal ($\alpha=0$); in this case, the quartet transition amplitude vanishes in the third order of perturbation theory, and this case needs a separate study.
All four contributions are distinguishable, since they are proportional to the product of three different interactions (e.g. $A_{k,lmn}\propto J_{kl}J_{km}J_{kn}$), and they cannot exactly compensate each other, except for the case of $\alpha=0$, that is beyond the scope of the present paper.
The quartet transition amplitude can also be calculated using Schrieffer-Wolff transformation, as in Ref. \cite{ab15MBLXY}, and it leads to the same result. 

This amplitude is very sensitive to the random energies, $\phi$, so it is convenient to replace it with the characteristic acting value. The acting value $\left<\left|V_{klmn} \right| \right>$ can be determined as the average absolute value of the amplitude in Eq. (\ref{eq:Vklmn}), similarly to that in the localization problem in the Bethe lattice \cite{AbouChacra73}. The logarithmic divergence at small denominators should be cut off at the critical energy,  $|\phi_{l}+\phi_{m}| \sim J_{lm}$, where the level repulsion becomes significant due to resonances. The average amplitude, $\left< \left| V_{klmn} \right| \right>$, can then be expressed as
\begin{align}
     \left< \left| V_{klmn} \right| \right> = \int_J^W  \frac{d^4\phi}{W^4} \left| V_{klmn}\right|.
    \end{align}
Average absolute value of the sum can, with the accuracy to the constant factor, be approximately replaced with the sum of absolute values of averages of each contribution leading to the sum over four contributing processes, lead by different spins $k$, $l$, $m$, $n$ respectively. The average amplitude can be represented as
    \begin{multline}
     \left< \left| V_{klmn} \right| \right> \sim \sum_{k,l,m,n} \frac{J_{kl}J_{km}J_{kn}}{ W^{2}} \left[ \ln \frac{W}{J_{kl}} \ln \frac{W}{J_{km}} +\right.\\\left.+ \ln \frac{W}{J_{km}}\ln \frac{W}{J_{kn}} +\ln \frac{W}{J_{kn}}\ln \frac{W}{J_{kl}}\right].
    \end{multline}
For the quartet made of spins separated by the intermediate distance $1 \leq R \leq N$ one can estimate the average transition amplitude as 
    \begin{align}
    V_* \sim W^{-2}\frac{J_0^3}{R^{3\alpha}} \ln^2 \left( \frac{R^\alpha}{J_0}W \right).
    \label{eq:vstar}
    \end{align}

In the case of small $\alpha$, the destructive interference leads to the reduction of the quartet transition amplitude, up to its complete vanishing for $\alpha=0$. This destructive interference may be responsible for the factor $\alpha$ in the numerical estimate of the critical disordering, as described in Sec. \ref{sec:Num}.

\subsection{Diagonal interaction} \label{diagonal}
The localization-delocalization transition is expected to happen when there are more than one spin-flipping resonances  \cite{ab15MBL,ab15MBLXY,Logan2018}; however, the mere existence of local resonances is not enough to establish chaotic behavior, because it is important that the diagonal coupling between separate resonances is strong compared to flipping amplitudes. In Ref. \cite{ab15MBLXY}, the diagonal interaction of spins has been estimated in the third order as an induced diagonal interaction, assisted by an additional spin $k$. Following the same logic for the Hamiltonian \eqref{eq:H}, one gets the diagonal corrections in the form $U_{ij}^{(3)}\sigma_i^z\sigma_j^z$, where the interaction $U_{ij}^{(3)}$ is defined as
\begin{align}
    U_{ij}^{(3)} = 4\sum_k \frac{\phi_i \phi_j J_{ij}J_{ik}J_{jk}}{\left(\phi_i^2 -\phi_k^2 \right)\left(\phi_j^2 -\phi_k^2 \right)},
\end{align}
which is given in the absence of the longitudinal field $B$ in Eq. \eqref{eq:H}, and generalization to the finite $B$ is straightforward. Since critical disordering $W_{c}$ increases limitlessly with the system size, one can ignore the field effect on statistics of interactions $U_*$ described below.

In case of $\alpha \leq 1$, this sum is determined by long distances $r_{ik} \sim r_{jk} \sim r_{ij} \sim N$, and by short distances $r_{ik} \sim 1$ or $r_{jk} \sim 1$, in the case of $\alpha>1$. The estimate for $U_*$ \cite{Raikh00Levy,ab15MBLXY} can then be written as
\begin{align}
   U_* \sim
  \begin{cases}
    J_0^3 W^{-2} N^{1-3\alpha}, & \text{$\alpha \leq 1$},\\
    J_0^3 W^{-2} N^{-2\alpha}, & \text{$\alpha > 1$}.
  \end{cases}  
  \label{eq:Ulong} 
\end{align}

The importance of the induced diagonal interaction and its influence on delocalized dynamics has been demonstrated not only for spin systems, but also for localization-chaos transition in the Fermi-Pasta-Ulam problem for vibrational dynamics in atomic chains \cite{BurinFPU2019}.

\subsection{Localization threshold}
The localization threshold is reached when the condition in Eq. \eqref{eq:resthr} is satisfied, meaning that the interaction energy between resonances exceeds their amplitude, $U_*>V_*$, and the number of resonances approaches unity within logarithmic accuracy. Depending on the case, the first or the second requirement is stronger and defines the threshold. Due to the different optimum choice of inter-spin distances in interacting quartets, the regimes of longer $\alpha\leq 1$ or shorter $\alpha >1$ range interactions are considered separately. We start our consideration with the most long-range case of $0<\alpha \leq 1$. 

\subsubsection{Case of $0<\alpha \leq 1$} \label{smallalpha}
For the case of $\alpha \leq 1$, the localization threshold can be determined by mapping the transverse field problem onto the Bethe lattice \cite{Burin2017} within the self-consistent theory of localization or the forward approximation. %The critical disordering can be found by comparing the contribution to the forward approximation with one, resulting in the condition obtained in Eq. \eqref{eq:Vklmn}, where $U_{*}$ stands for the interaction between quartets obtained in \eqref{eq:Ulong}. The quartet resonance amplitude $\left<\left|V_{klmn} \right| \right>$ in this case is determined by the largest distances of order of $N$, and can be estimated up to a factor from Eq. \eqref{eq:vstar} setting $\ln N^\alpha W/J_0 \sim \ln N$ which results in
The critical disordering is determined from Eq. \eqref{eq:Vklmn}, where the largest contribution is made by quartets of spins at distances $R \sim N$ from each other. Setting $R \sim N $ in Eq. \eqref{eq:vstar}, and neglecting small corrections to the logarithmic factor, we get 
\begin{align}
    V(N) = J_0^3 W^{-2} N^{-3\alpha} \ln^2 N.
\end{align}
%The quartet resonance amplitude is clearly smaller than the characteristic diagonal interaction of resonances at the maximum distance given by Eq. \eqref{eq:Ulong}, which justifies Eq. \eqref{eq:Vklmn}.

%The interaction $U(N)$ given by Eq. \eqref{eq:Ulong}  yields $\ln U(N)/V(N) \sim \ln N$; as we will see below there is no such factor for $\alpha>1$. Summing up all $N^4$ possible quartets, the condition in Eq. \eqref{eq:Vklmn} defines the critical disordering in the form 
The quartet resonance amplitude, $V(N)$, is clearly smaller than the characteristic diagonal interaction, $U_{*} \sim J_{0}^3W^{-2}N^{1-3\alpha}$, that allows us to use Eqs. \eqref{eq:resthr} and \eqref{eq:Vklmn} to estimate the localization threshold. Setting $\ln(U_{*}/V) \approx \ln N$ and $N_* \sim N^4$, one can express the localization threshold as
\begin{align}
    W_c \propto J_0N^{4/3-\alpha} \ln N.
    \label{eq:Wc}
\end{align}

\subsubsection{Case of $\alpha>1$} \label{largealpha}

If $\alpha > 1$, the density of resonant quartets of the typical size $R <N$ scales as $R^{3(1-\alpha)}$, which means that it increases as the size of the quartet decreases. Consequently, the delocalization can be associated with quartets of the size $R<N$, which are characterized by the transition amplitude $V(R) \sim J_{0}^3 W^{-2}R^{-3\alpha}\ln^2(N)$ (cf. Eq. \eqref{eq:vstar}). These quartets lead to delocalization when few of them are formed if their amplitude $V(R)$ is less or equal to their longitudinal interaction $U_* \sim J_{0}^3W^{-2}N^{-2\alpha}$, as defined in Eq. \eqref{eq:Ulong}. Setting $U_* \sim V(R)$ and the number of quartets to unity, we obtain the delocalization threshold in the form
\begin{align}
   W_c \propto J_0 N^{1-2\alpha/3} \ln^{2/3}N.
\end{align}
In this case, the logarithmic dependence is weaker, than in the case of $\alpha<1$, because $U_{*} \sim V_{*}$ so that the logarithmic factor in Eq. \eqref{eq:resthr} is of order of unity. The result is applicable only for $\alpha < 3/2$ where the critical disordering increases with the system size. The interaction power law exponent constraint $\alpha < 3/2$ agrees with Ref. \cite{Burin2017}, where interacting resonances of spin pairs have been considered. The scaling exponent of the critical disordering is different from that in Ref. \cite{Burin2017} by the factor $2(1+\alpha)/3$. Below, we argue that the consideration of more complicated spin excitations (sextets or more) does not modify this criterion. 

\subsubsection{Multi-spin clusters}
The resonant quartets, rather than the resonant pairs, determine the localization transition in the problem under consideration, because the longitudinal interaction of pairs is insufficiently strong to suppress the destructive interference of their transitions (cf. Ref. \cite{Burin2017}); on the other hand, the more complicated spin excitations possess the smaller probability of resonance, and, therefore, we do not expect them to modify the critical disordering estimated for quartets. Indeed, for $n$-spin resonances, one can estimate their amplitude using $(n-1)^{st}$ order of perturbation theory (e. g. simultaneous transitions of the first spin with $n-1$ other spins) as
\begin{align}
V_{n} \sim N^{-\alpha (n-1)}J_{0}^{n-1}/W^{n-2}.
\end{align}
We consider here only the case $\alpha \leq 1$, where all characteristic distances are of order of $N$; the case $\alpha>1$ can be treated similarly. The logarithmic factors are ignored for the sake of simplicity. The number of resonances can be estimated by multiplying the probability of individual resonances, $V_{n}/W$, by the number of $n$-spin combinations, $N^{n}$, which yields 
\begin{align}
N_{res} \sim N^{(1-\alpha) (n-1)+1}J_{0}^{n-1}/W^{n-1}.
\end{align}
Setting $N_{res} \sim 1$ we get the critical disordering estimate as
\begin{align}
W_{c} \sim J_{0}N^{n/(n-1)-\alpha},
\end{align}
which obviously has a maximum at $n=4$ corresponding to quartets (remember that $n$ must be even). Therefore, we believe that the localization breakdown is determined by quartets, which is confirmed by the numerical results reported in Sec. \ref{sec:Num}.

\subsubsection{Summary of analytic predictions}
Both results for the localization threshold obtained in two different regimes of large (Sec. \ref{largealpha}) and small (Sec. \ref{smallalpha}) power law exponent $\alpha$ can be resumed as following
\begin{eqnarray}
W_c =c_{\alpha}
\begin{cases}
    J_0 N^{4/3-\alpha} \ln N, & \text{$0 < \alpha \leq 1$},\\
    J_0 N^{1-2\alpha/3} \ln^{2/3} N, & \text{$1 < \alpha < \frac{3}{2}$}, 
  \end{cases}  
  \label{eq:Case4} 
\end{eqnarray}
where $c_{\alpha}$ is a constant numerical factor determined below in Sec. \ref{sec:Num} by fitting the numerical results with our analytical expression.

The result in Eq. \eqref{eq:Case4} cannot be extended to the case of $\alpha=3/2$, since the dependence $W_{c} \propto \ln^{2/3}N$ relies on the factor $\ln^{2/3}(W_{c}/J)$, while $W_{c}$ does not show the power-law dependence of $N$ at $\alpha =3/2$; therefore, we don't have any reasonable prediction for the localization threshold scaling in this crossover regime. The numerical analysis of Sec. \ref{sec:Num} is also inconclusive in this case. 

The obtained scalings of the critical disordering that correspond to the localization threshold \eqref{eq:Case4} should be observable in the dependencies of the level statistics and spin-spin correlation functions. These parameters will be studied numerically in Sec. \ref{sec:Num} using exact diagonalization of the problem in Eq. \eqref{eq:H}, and it will be shown that the numerical findings are consistent with the analytic predictions of the present section.

\subsection{Finite $W_{c}$ in the thermodynamic limit}
\label{sec:thermlimresc}

Since the critical disordering, $W_c$, in Eq. (\ref{eq:Case4}) becomes infinite as the number of spins, $N$, approaches infinity, there is no localization transition in the thermodynamic limit for the system described by Eq. (\ref{eq:H}). Following the spin glass model  \cite{Sherrington75} and Rosenzweig-Porter random matrix model \cite{RosenzweigPorterOrig60,KravtsovRozenPort2015}, one can rescale the spin-spin coupling strength in the Hamiltonian Eq. (\ref{eq:H}) as (cf. Eq. (\ref{eq:Case4}) in the large $N$ limit)  
\begin{eqnarray}
    \tilde{J}_{0} \rightarrow  \frac{J_0}{c_{\alpha}N^{1/3+\xi(1-\alpha)}\ln^{\xi}N}, \nonumber \\ \xi= 
    \begin{cases}
        1,  & 0 < \alpha \leq  1, \\
        \frac{2}{3}, & 1< \alpha < \frac{3}{2}.
    \end{cases}.
    \label{eq:Resc}
\end{eqnarray}
After this rescaling the critical disordering approaches the size independent limit $W_{c}=J_{0}$, while the transverse interaction becomes too weak to enter any thermodynamic parameter, yet it is sufficient to bring the system to the chaotic state.

Comparing the mentioned rescaling to the Katz prescription as in Refs \cite{Hauke15MBLLongRange,Zakrzewski19}, it can be noticed that the latter is not strong enough to make the critical disordering size independent in the limit of large N. A weak size dependence of critical disordering, $W_{c} \propto N^{1/3}$, remains after applying Katz prescription. Ref. \cite{Zakrzewski19} illustrates a significant effect of Katz prescription on delocalization in the case of long-range diagonal interaction; yet, similarly to the previous consideration we expect delocalization even in that regime with the scaling preliminary estimated as $W_c \propto N^{1/2}$. The more accurate analysis of that system should be performed separately.  

A system described by Eq. (\ref{eq:H}) with the interaction constant, $\tilde{J}_0$, redefined according to Eq. (\ref{eq:Resc}) is stable with respect to the formation of chaotic spots \cite{HotSpotLuitz2017,Huveneers17BrekDwnLoc} of several neighboring spins with reduced random fields, $|\phi_i| \sim \tilde{J}_{0} \ll  W \sim J_{0}$. Since chaotic spots are formed by rare fluctuations of random energy making several adjacent spins chaotic, the random energy for such a spot should be comparable to the coupling strength decreasing as $N^{-\eta}$, where $\eta=1/3+\xi(1-\alpha)$ (see Eq. (\ref{eq:Resc})). Logarithmic dependencies can be omitted here  as less significant compared to the power laws. 
 
The probability to create a chaotic spot of $k$ spins scales as $P_{c}(k) \sim N^{-\eta k}$ with the maximum number of spins limited by the constraint $k<1/\eta$. Chaotic spots containing more spins can be neglected since the total probability to form such spots, $NP_{c}(k)$, vanishes in the thermodynamic limit of infinite $N$. Surrounding spins that can exchange energy with the chaotic spot of $k$ spins should have random fields, $\phi_i$, not exceeding the maximum spot energy $\sqrt{k} N^{-\eta} \ll 1$. The distance to closest spin satisfying this condition is $r \sim N^{\eta}/\sqrt{k}$ and the interaction with it can be estimated as $\tilde{J}_{0}r^{-\alpha} 2^{-k/2} \sim N^{-\eta (\alpha+1)}2^{-k/2}$, cf. Ref.  \cite{Huveneers17BrekDwnLoc}. 
To add an external spin to the chaotic spot this interaction should exceed the level splitting within the spot,  which can be estimated as $\delta \sim N^{-\eta}2^{-k}$. 
Since the maximum number of spins in the chaotic spot is finite, the spot-spin resonance condition, $2^{k/2} > N^{\eta\alpha}$, cannot be satisfied in the thermodynamic limit of infinite $N$.

%It can be summarized that the localization-delocalization transitions follows two different mechanisms for $\alpha < 3/2$ and $\alpha > 3/2$. In the first case of violated dimensional constraint, a slightly modified Bethe-lattice scenario \cite{Burin2017} is applicable, while in the second case, the transition is determined by rare fluctuations like hot spots. 

\subsection{Transition at small $W$ for $B=0$}

As it was noticed in the introduction, for $B=0$ there is also a localization transition at $W \ll J_0$; this transition can be described following Ref. \cite{Maksymov2017} and the localization takes place at $W < W_{l}$, defined as:
\begin{eqnarray}
W_{l} \sim
\begin{cases}
   J_0 N^{-(\alpha+1/2)} & 0 < \alpha\leq \frac{1}{2},\\    
   J_0 N^{-1} & \frac{1}{2}< \alpha < 1,\\
   J_0 N^{-(2-\alpha)} & 1\leq  \alpha < 2. 
\end{cases}
\label{eq:res}
\end{eqnarray}
Since this transition is already characterized in Ref. \cite{Maksymov2017}, we do not focus on that regime. In the case of a finite field $B$ in Eq. \eqref{eq:H}, considered in the experiment \cite{Monro16}, this transition does not take place. 

\section{Numerical studies}
\label{sec:Num}

The localization transition in the system described by Eq. \eqref{eq:H} is investigated using Hamming distance \cite{AltshullerGefen97,Monro16,Hauke15MBLLongRange}  and level statistics, expressed in terms of the average minimum ratio of adjacent energy differences \cite{OganesyanHuse07}. The normalized Hamming distance between the initial and final states can be directly measured experimentally. The Hamming distance between two Ising states is measured as the number of flips required to change one state into the other. The normalized Hamming distance is given as a ratio of the number of spin flips to the total number of spins. At certain time, $t$, and all spins in initial states determined by the sequence $\sigma_i^z(0)$, the normalized Hamming distance can be expressed using the initial state $\left| \psi_0 \right>$ as
\begin{align}
D(t)=\frac{1}{2N}\sum_i \left< \psi_0 \right| \left[ \sigma_i^z(0) - \sigma_i^z(t) \right] \sigma_i^z(0)\left| \psi_0 \right>.
\label{eq:Di}
\end{align}
In the fully delocalized regime in the long time limit, the normalized Hamming distance approaches its maximum $1/2$, due to thermalization, while in the fully localized state it remains zero.

The level statistics is represented by the averaged ratio of minimum to maximum differences between successive eigenenergies of the system
\begin{align}
    \left< r \right> = \left< \frac{\min\left(\Delta_n,\Delta_{n+1}\right)}{\max\left(\Delta_n,\Delta_{n+1}\right)} \right>_n.
    \label{eq:lst} 
\end{align}
The localized regime obeys Poisson statistics for $\Delta_n$, and is characterized by $\left< r \right> \approx 0.3863$, while the delocalized regime is known to obey the Wigner-Dyson statistics with $\left< r \right> \approx 0.5307$ \cite{OganesyanHuse07}. To calculate the level statistics numerically one has to find eigenvalues of the system Hamiltonian and average them over different disorder realization. For the Hamming distance, the eigenstates of the system also need to be found.

All calculations were performed in MATLAB \cite{Bridges}. Eigenenergies and eigenstates of the Hamiltonian \eqref{eq:H} were found through exact diagonalization, and averaged over 2000 realizations of random potentials for every disorder. %Averaging for $N<16$ is done over 1000 realizations of disorder. For larger $N$ the level statistics is averaged over bigger number of states with energies around zero. 

%Hamming distance and level statistics were studied in the range of $8\leq N \leq 16$ with the transverse field $B=0$ and $B=4J_0$, as in Ref. \cite{Monro16}. For the case of $B=4J_0$ the set of power-law interaction exponents $\alpha=0.25,\,0.5,\,0.85,\,1,\,1.15,\,1.25,$ and $1.5$ was studied, while for the case of $B=0$ we studied $\alpha=0.25,\,0.5,\,1$ and $1.5$.

Hamming distance and level statistics were studied in the range of $8\leq N \leq 16$ with a transverse field, $B$. The results below are given for the case of $B=4J_{0}$ (as in Ref. \cite{Monro16}) and power-law interaction exponents $\alpha=0.25,\,0.5,\,0.85,\,1,\,1.15,\,1.25,$ and $1.5$. We also studied the case of $B=0$, for which the results were quite similar, except for the domain around $W=0$ described by Eq. \eqref{eq:res}. Since this work is focused on the transition at large $W$, we present only the results for $B=4J_{0}$. 
%For three selected values of $\alpha=0.5,1,$ and $1.5$, an additional set of level statistics was calculated for $B=0$, to compare to the finite transverse field data.
For all calculations, the coupling constant, $J_0$, was set to unity.

Numerical results are to be compared with the theory predictions expressed by Eq. \eqref{eq:Case4}. To account for the discrete effects at small sizes, the power-law dependence of $N$ in Eq. (\ref{eq:Case4}) is substituted with the finite sum, similarly to Ref.  \cite{ab15MBLXY}, as
\begin{align}
    I(N,\beta) = \sum_{R=4}^{N} R^{-\beta} \approx \frac{N^{1-\beta}}{1-\beta}, 
\end{align}
where $1-\beta = \eta = 1/3+\xi(1-\alpha)$ (see Eq. \eqref{eq:Resc}) chosen to satisfy the predicted scaling of Eq. \eqref{eq:Case4} in the limit of infinite $N$. 
The critical disordering can then be given in the form consistent with Eq. (\ref{eq:Case4}) as
\begin{align}
    W_c = 
c_{\alpha} \eta I(N,1-\eta) \ln^\xi N
  \label{eq:nWc} 
\end{align}
The numerical optimization of data collapses for Hamming distance and level statistics described below leads to the accurate definition of the proportionality factors, $c_{\alpha}$, that can be expressed as
\begin{align}
c_{\alpha}\approx \begin{cases}
    \frac{1.37\alpha}{4/3-\alpha}, & \text{$0 < \alpha \leq 1$},\\
    \frac{\alpha}{1-2\alpha/3}, & \text{$1 < \alpha < \frac{3}{2}$}.
  \end{cases}  
  \label{eq:Calpha} 
\end{align}

Hamming distance provides better information about delocalization transition, because it is less sensitive to the symmetry and integrability of the problem in the limit of small disordering, showing the most pure scaling for the transition at large $W$; therefore, we begin our consideration with the Hamming distance. The level statistics is also very important, since the observation of the Wigner-Dyson statistics gives the best, basis independent evidence for the chaotic behavior; therefore, we analyze level statistics in Sec. \ref{sec:num:lst}, and demonstrate that the obtained behaviors are consistent with those found using the Hamming distances. 

\subsection{Hamming distance}
The correlation functions of spins determine the Hamming distance between initial spin state, chosen as anti-ferromagnetic $\text{N\'{e}el}$ state \cite{Monro16}, and the state at time $t$. This distance is defined in Eq. \eqref{eq:Di}, which can be rewritten in terms of the system eigenstate basis as 
\begin{align}
    D(t) = \frac{1}{2} - \frac{1}{2N} \sum_{nm} \left< \psi_0 \right| \left. n \right>^2 |\sigma^z_{mn}|^2  e^{-it(E_n-E_m)/\hbar},
\end{align}
where summation goes over all eigenstates $\left|n\right>$ and $\left|m\right>$, having energies $E_n$ and $E_m$, correspondingly. In the limit $t\rightarrow \infty$ the oscillating terms with $n \neq m$ can be neglected and the last expression takes the form:
\begin{align}
    D(\infty) = \frac{1}{2} - \frac{1}{2N} \sum_{n} \left< \psi_0 \right| \left. n \right>^2 |\sigma^z_{nn}|^2.
\end{align}
Below, we start with comparison of numerical results with experimental observations in Ref. \cite{Monro16}, then perform data rescaling to analyze the dependence of localization threshold and transition width on the system size, and then consider the dependence of critical disordering on the power-law interaction exponent $\alpha$.

\subsubsection{Numerical results for infinite time}
Comparison to the experimental results \cite{Monro16} at large finite time shows that numerical results for the Hamming distance give a slightly larger value compared to the experimental data. 
\begin{figure}[h!]
\centering
\includegraphics[width=0.8\columnwidth]{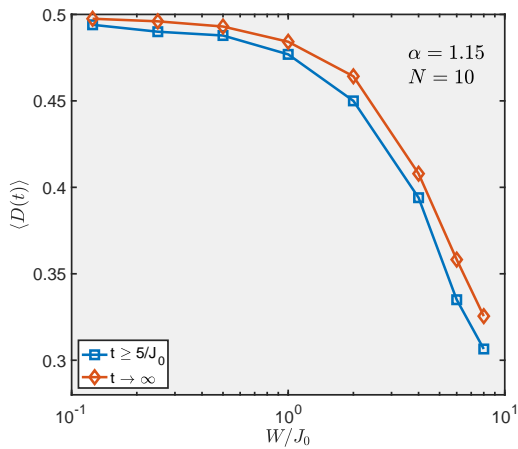}
\caption{\small The normalized Hamming distance according to Eq. \eqref{eq:Di} (red line) and extracted from the experiment \cite{Monro16} for finite time $t \geq 5/J_0$ (blue line) for $\alpha=1.15$ and $N=10$.}
\label{fig:Dt}
\end{figure}
As it has been mentioned above, the normalized Hamming distance grows with time, due to the thermalization process to reach $1/2$ for the fully delocalized regime at infinite time. Since the experimental data provided are not at infinite time, the measured Hamming distance is expected to be a bit less than the numerically calculated, which can be observed on the $\text{Fig. \ref{fig:Dt}}$.

\subsubsection{Scaling with the system size}
\begin{figure}[h!]
\centering
\includegraphics[width=.8\linewidth]{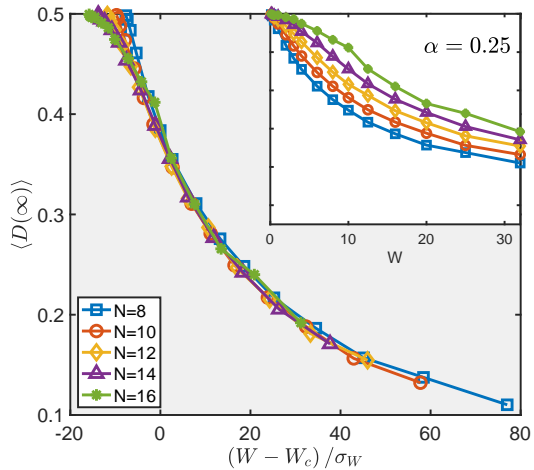}
\caption{\small The normalized Hamming distance at infinite time vs disordering rescaled as $(W-W_{c})/\sigma_{W}$ (main plot) with original data (inset) for $8\leq N \leq 16$ and $\alpha=0.25$.}
\label{fig:ScaleN025}
\end{figure}

\begin{figure}[h!]
\centering
\includegraphics[width=.8\linewidth]{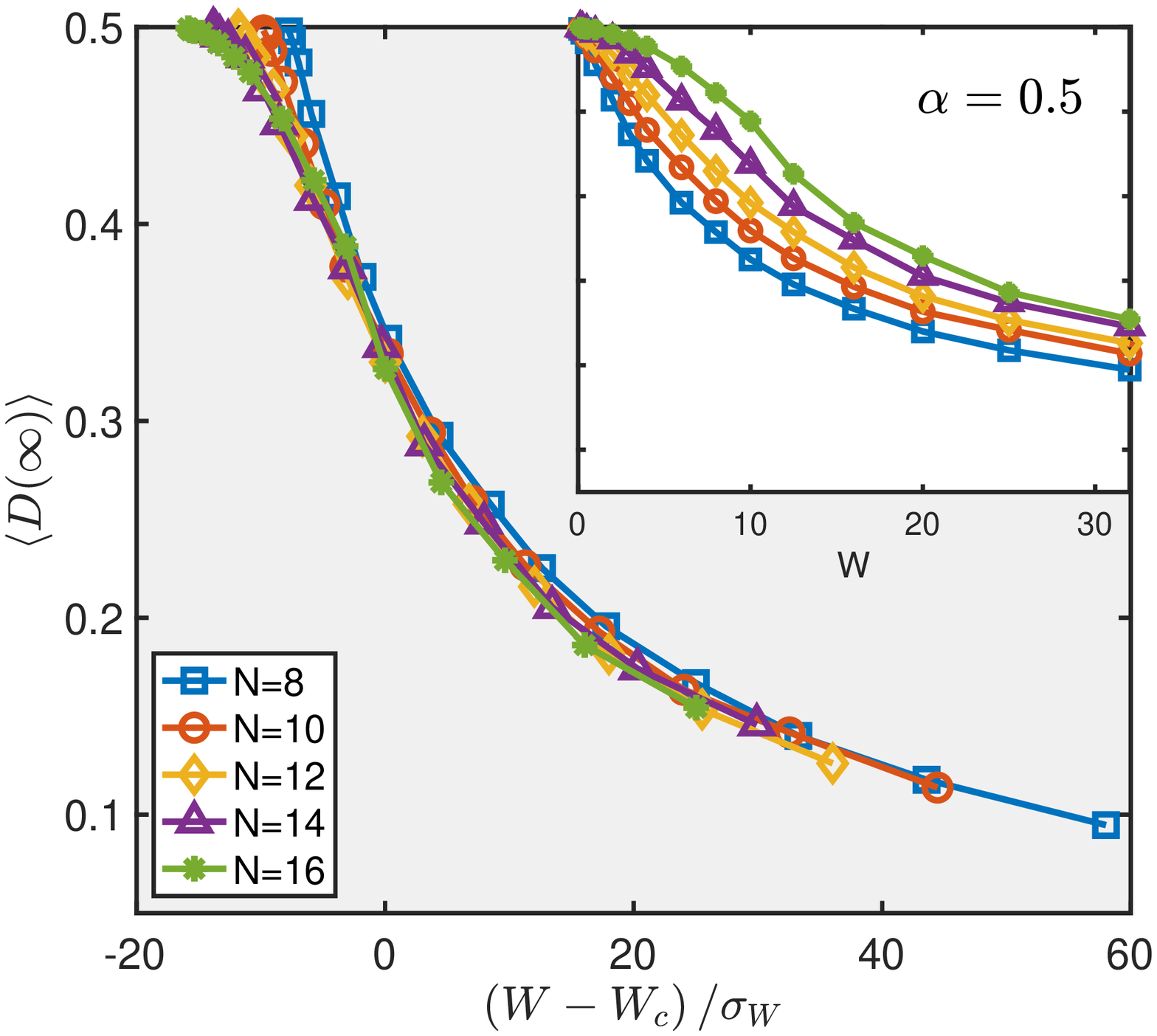}
\caption{\small The normalized Hamming distance at infinite time vs disordering rescaled as $(W-W_{c})/\sigma_{W}$ (main plot) with original data (inset) for $8\leq N \leq 16$ and $\alpha=0.5$.}
\label{fig:ScaleN05}
\end{figure}

\begin{figure}[h!]
\centering
\includegraphics[width=.8\linewidth]{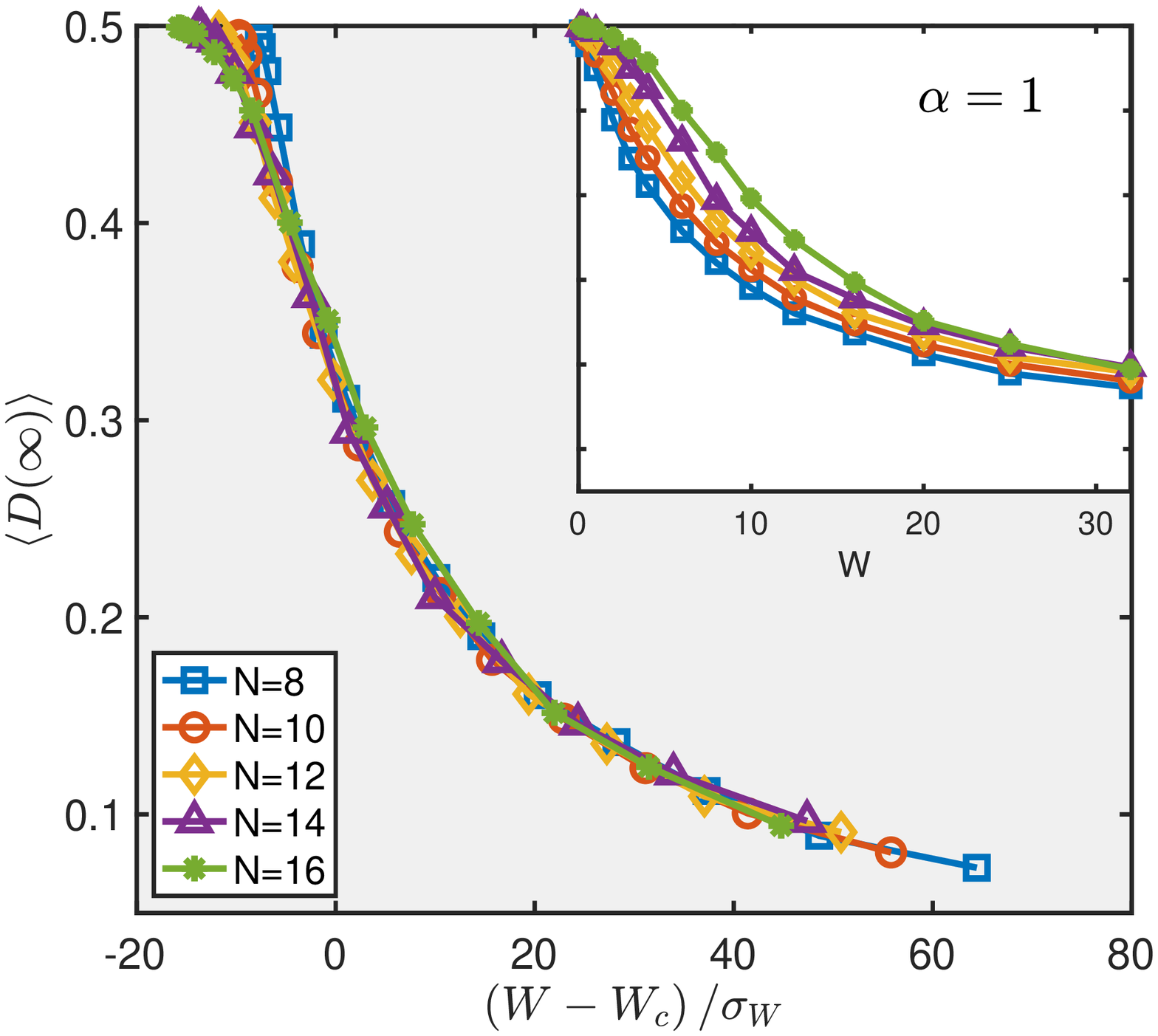}
\caption{\small The normalized Hamming distance at infinite time vs disordering rescaled as $(W-W_{c})/\sigma_{W}$ (main plot) with original data (inset) for $8\leq N \leq 16$ and $\alpha=1$.}
\label{fig:ScaleN1}
\end{figure}

\begin{figure}[h!]
\centering
\includegraphics[width=.8\linewidth]{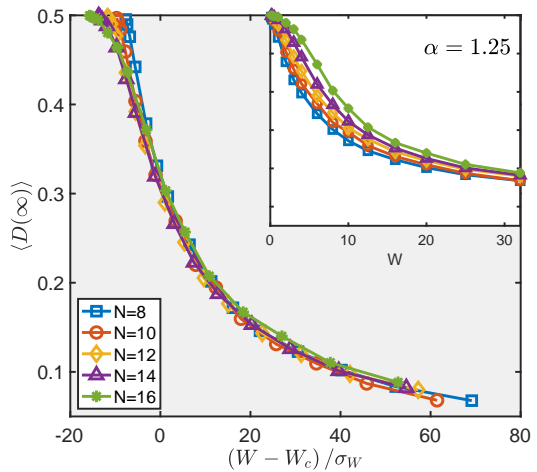}
\caption{\small The normalized Hamming distance at infinite time vs disordering rescaled as $(W-W_{c})/\sigma_{W}$ (main plot) with original data (inset) for $8\leq N \leq 16$ and $\alpha=1.25$.}
\label{fig:ScaleN125}
\end{figure}

To analyze the data, we plot Hamming distances vs. disordering rescaled as $(W-W_{c})/\sigma_{W}$, where the critical disordering $W_c$ is expressed from the equation \eqref{eq:nWc} with $c_{\alpha}$  as in Eq. (\ref{eq:Calpha}) and the transition width $\sigma_{W}$ chosen as
\begin{align} 
\sigma_{W}=\frac{W_c}{N}.
\end{align}
There, rescaled graphs collapse onto one curve, as can be seen on \cref{fig:ScaleN025,fig:ScaleN05,fig:ScaleN1,fig:ScaleN125} for selected $\alpha=0.25,\,0.5,\,1,$ and $1.25$, respectively.

Bi-parametric optimization has been performed for both threshold disordering coefficient $k$ and rescaling exponent $\beta$ \cite{Pollmann2014}. The scaling of transition width as $W_{c}/N$ was found based on numerical analysis of the best data collapse. This result can be interpreted using the following estimate of the minimum transition width: the probability for each random potential to stay within the range $-(W-\delta W)/2 < \phi_i < (W-\delta W)/2$ is $p(W) = 1-\frac{\delta W}{W}$; then, the probability of all random potentials to fall into the same range is
\begin{align}
p_N = \left(1-\frac{\delta W}{W}\right)^N \approx 1-\frac{N}{W}\delta W,
\end{align}
which means that two realizations of random potentials differing by less than $W/N$ are not distinguishable; that yields the estimate $\sigma_{W} = W_{c}/N$, consistent with the results below.

%\begin{figure}[h!]\centering\includegraphics[width=.8\linewidth]{HA085.eps}\caption{\small The normalized Hamming distance at infinite time rescaled as $(W-W_{c})/\sigma_{W}$ for $N=8,10,12,14$ and $\alpha=0.85$ (main plot) with original data without rescaling (inset).}\label{fig:ScaleN085}\end{figure}

%\begin{figure}[h!]\centering\includegraphics[width=.8\linewidth]{HA115.eps}\caption{\small The normalized Hamming distance at infinite time rescaled as $(W-W_{c})/\sigma_{W}$ for $N=8,10,12,14$ and $\alpha=1.15$ (main plot) with original data without rescaling (inset).}\label{fig:ScaleN115}\end{figure}

%It can be seen from the graphs (Fig. \ref{fig:HammA16}) that there is a good collapse for of all $N$ curves. Although while taking a closer look at subplot with original dependencies one can notice that the transition width is increasing with the system size. That seems to contradict the expected $\sigma_W \propto N^{1/3-\alpha}$ which should reduce for $\alpha>1/3$. Such behavior is a result of finite size effects since in the considered range of $N$ the logarithmic factor behaves almost as $N^{2/3}$.

\begin{figure}[h!]
\centering
\includegraphics[width=.8\linewidth]{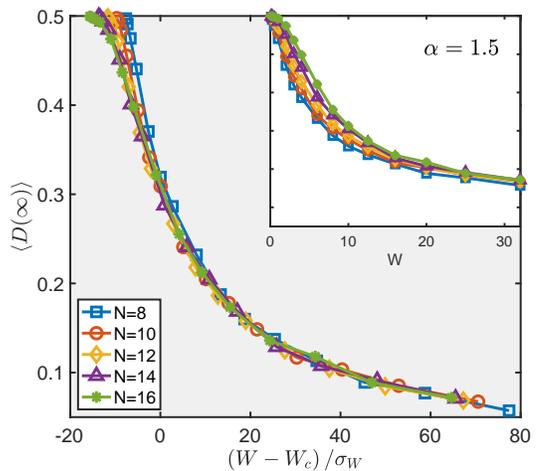}
\caption{\small The normalized Hamming distance at infinite time vs disordering rescaled as $(W-W_{c})/\sigma_{W}$ (main plot) with original data (inset) for $8\leq N \leq 16$ and $\alpha=3/2$.}
\label{fig:ScaleN15}
\end{figure}

\begin{figure}[h!]
\centering
\includegraphics[width=0.8\columnwidth]{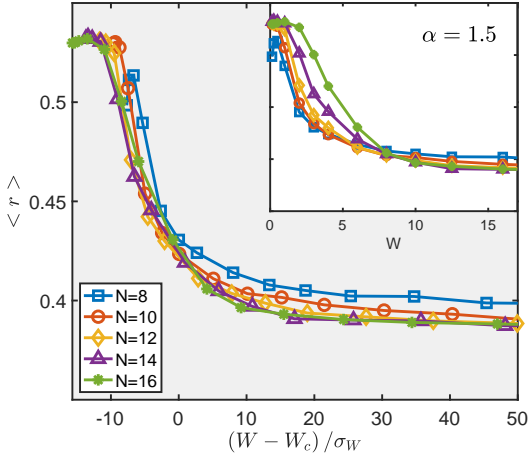}
\caption{\small The level statistics vs disordering rescaled as $(W-W_{c})/\sigma_{W}$ (main plot) with original data (inset) for $8\leq N \leq 16$ and $\alpha=3/2$.}
\label{fig:Stat15}
\end{figure}

The data scaling for the threshold $\alpha=3/2$ (Fig. \ref{fig:ScaleN15}) are surprisingly consistent with Eq. \eqref{eq:nWc}, in spite of the irrelevance of the derivation as explained after Eq. \eqref{eq:nWc}. The observed logarithmic dependence can have different origins, including, for instance, ergodic spots \cite{Huveneers17BrekDwnLoc,Mirlin16Fractal}; this can be the reason for the special behavior of level statistics in this case (see Fig. \ref{fig:Stat15}).

\subsubsection{Scaling with the power-law exponent}
\begin{figure}[h!]
\centering
\includegraphics[width=0.8\columnwidth]{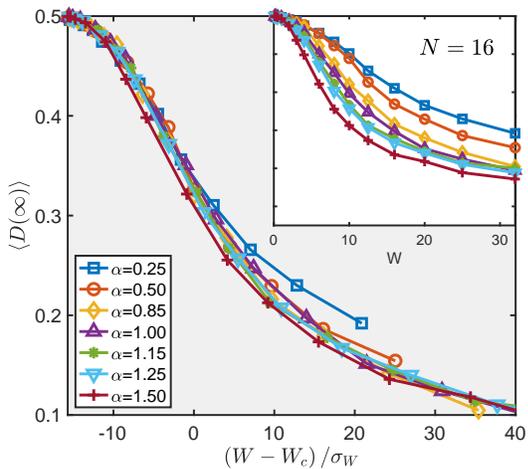}
\caption{\small The normalized Hamming distance at infinite time vs disordering rescaled as $(W-W_{c})/\sigma_{W}$ (main plot) with original data (inset) for $0.25 \leq \alpha \leq 3/2$ and $N=16$.}
\label{fig:HammA16}
\end{figure}

As it is clear from \cref{fig:ScaleN025,fig:ScaleN05,fig:ScaleN1,fig:ScaleN125}, the fit of the critical disordering by Eq. (\ref{eq:nWc}) gives a very good data collapse for different power-law interaction exponents, $\alpha$. Placing all data for different numbers of spins, $N$, and interaction exponents, $\alpha$, onto one graph, it can be seen that all data for $0.5<\alpha<1.5$ can be represented by a single curve reasonably well (see Fig. \ref{fig:HammA16}).  The deviations for small $\alpha$ can be due to stronger correlations between different interactions vanishing at $\alpha=0$. This similarity supports the expectation of Sec. \ref{sec:thermlimresc} that MBL transitions at violated dimensional constraint are similar to the localization transition in the Bethe lattice and, therefore, they show similar behaviors.

\subsection{Level statistics}
\label{sec:num:lst}

The level statistics is calculated according to Eq. \eqref{eq:lst}, where averaging is done in a narrow ($n \approx 100$) range of eigenstates, with energies around zero, which corresponds to the infinite temperature. Generally speaking, since there is no restriction in participating eigenstates when the Hamming distance is calculated, the results of two scalings don't have to match.

%For both finite field, $B=4J_0$, and zero field, $B=0$, the limit of zero disordering $W$ should lead to the deviations of the level statistics from the Wigner-Dyson behavior. In the latter case of a zero field, the localization domain is expected to be formed at small, but finite $W$ (see Eq. (25)). In the former case of a finite $B$, the system has a reflection symmetry at $W=0$ that leads to the separation of states into symmetric and antisymmetric (see Ref. \cite{Maksymov2017}), and a breakdown of Wigner-Dyson statistics in spite of delocalization. Consequently, in the case of finite field $B=4J_0$, the scaling of the transition works much better for all power-law interaction exponents, compared to the case of $B=0$, because of the finite size effects, which should be less significant at larger system sizes; below, we restrict our consideration to the case of $B=4J_0$, although cases of $\alpha=0.25,0.5,$ and $1$ were also studied for $B=0$ and they show a reasonable collapse after rescaling. 

\begin{figure}[h!]
\centering
\includegraphics[width=0.8\columnwidth]{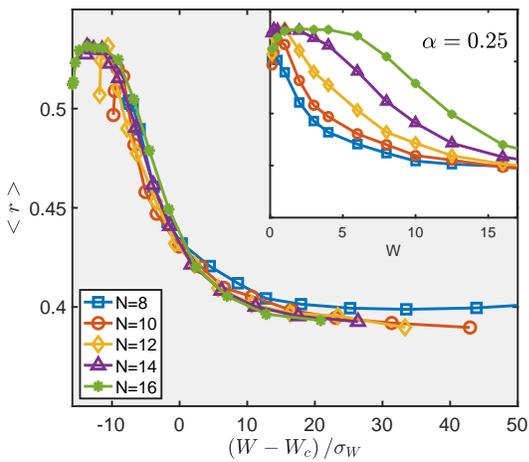}
\caption{\small The level statistics vs disordering rescaled as $(W-W_{c})/\sigma_{W}$ (main plot) with original data (inset) for $8\leq N \leq 16$ and $\alpha=0.25$.}
\label{fig:Stat025}
\end{figure}

\subsubsection{Scaling with the system size}

For the case of interest, $B=4J_0$, the level statistics was rescaled following the same law, $(W-W_{c})/\sigma_{W}$, and keeping the coefficients found for the corresponding cases of Hamming distance (\cref{fig:ScaleN025,fig:ScaleN05,fig:ScaleN1,fig:ScaleN125}). %For larger system sizes, there is a much better fit, and no second transition for small disorderings, as the symmetry is broken by the transverse field. As it can be seen, $N=8$ chain appears to be too small, and the level statistics shows it much better than the Hamming distance. (Fig. \ref{fig:Stat05})
The fit is gets better as the system size increases, because of rapid narrowing of the weak disorder domain, affected by symmetry at $W=0$. 

\begin{figure}[h!]
\centering
\includegraphics[width=0.8\columnwidth]{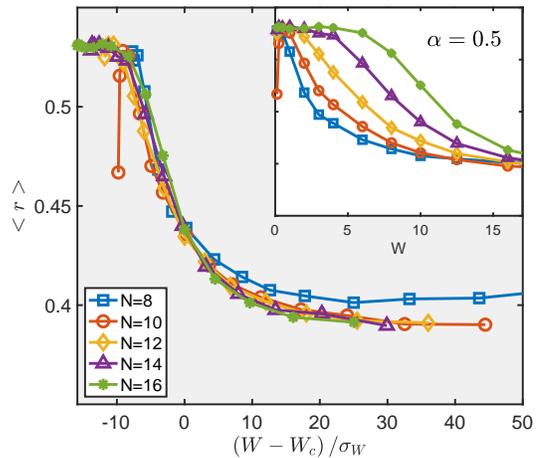}
\caption{\small The level statistics vs disordering rescaled as $(W-W_{c})/\sigma_{W}$ (main plot) with original data (inset) for $8\leq N \leq 16$ and $\alpha=0.5$.}
\label{fig:Stat05}
\end{figure}

%\begin{figure}[h!]\centering\includegraphics[width=0.8\columnwidth]{SA085.eps}\caption{\small The rescaled level statistics (main plot) and the original level statistics (inset) for $\alpha=0.85$ and $B=4$.}\label{fig:Stat085}\end{figure}

\begin{figure}[h!]
\centering
\includegraphics[width=0.8\columnwidth]{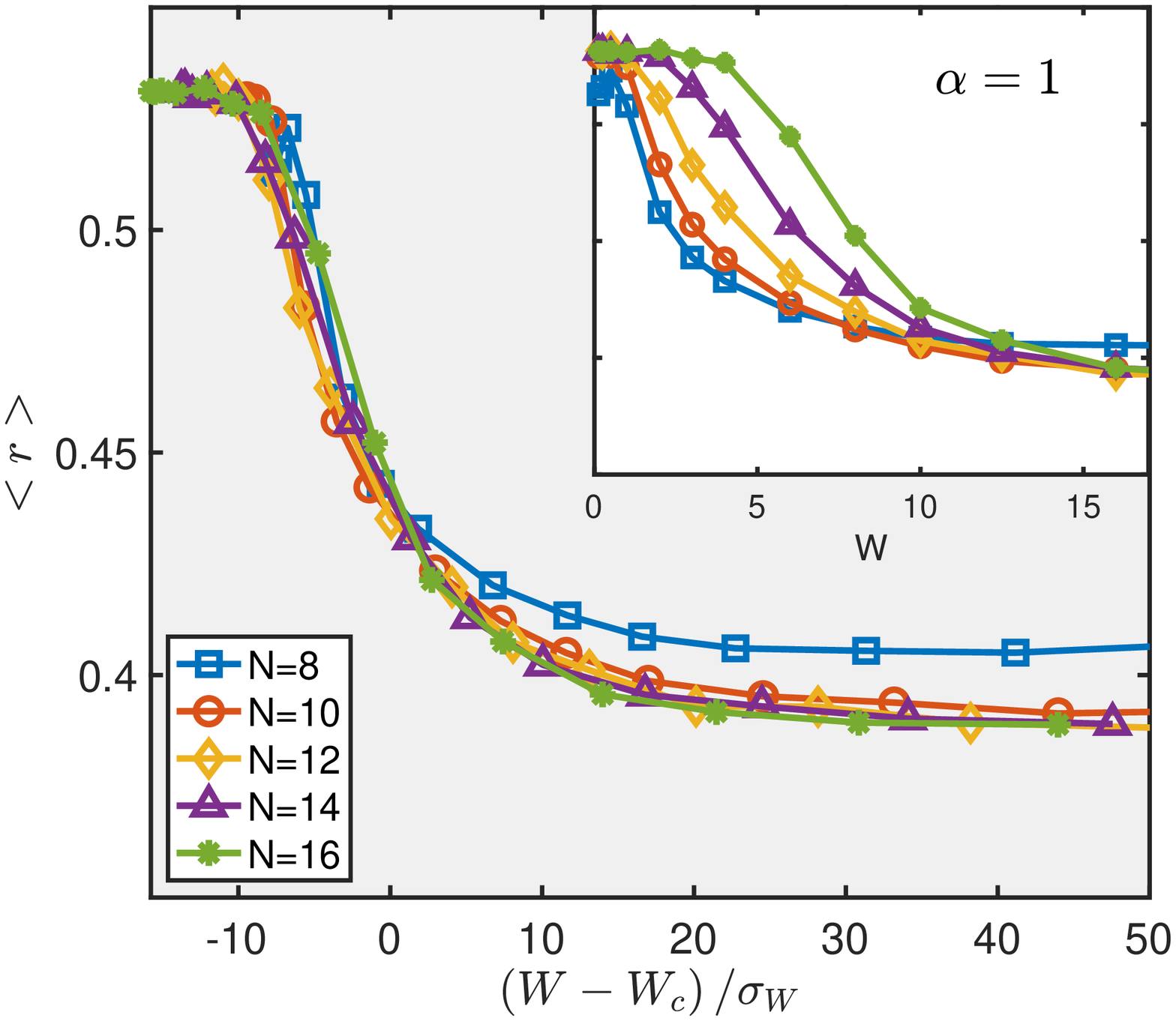}
\caption{\small The level statistics vs disordering rescaled as $(W-W_{c})/\sigma_{W}$ (main plot) with original data (inset) for $8\leq N \leq 16$ and $\alpha=1$.}
\label{fig:Stat1}
\end{figure}

%\begin{figure}[h!]\centering\includegraphics[width=0.8\columnwidth]{SA115.eps}\caption{\small The rescaled level statistics (main plot) and the original level statistics (inset) for $\alpha=1.15$ and $B=4$.}\label{fig:Stat115}\end{figure}

\begin{figure}[h!]
\centering
\includegraphics[width=0.8\columnwidth]{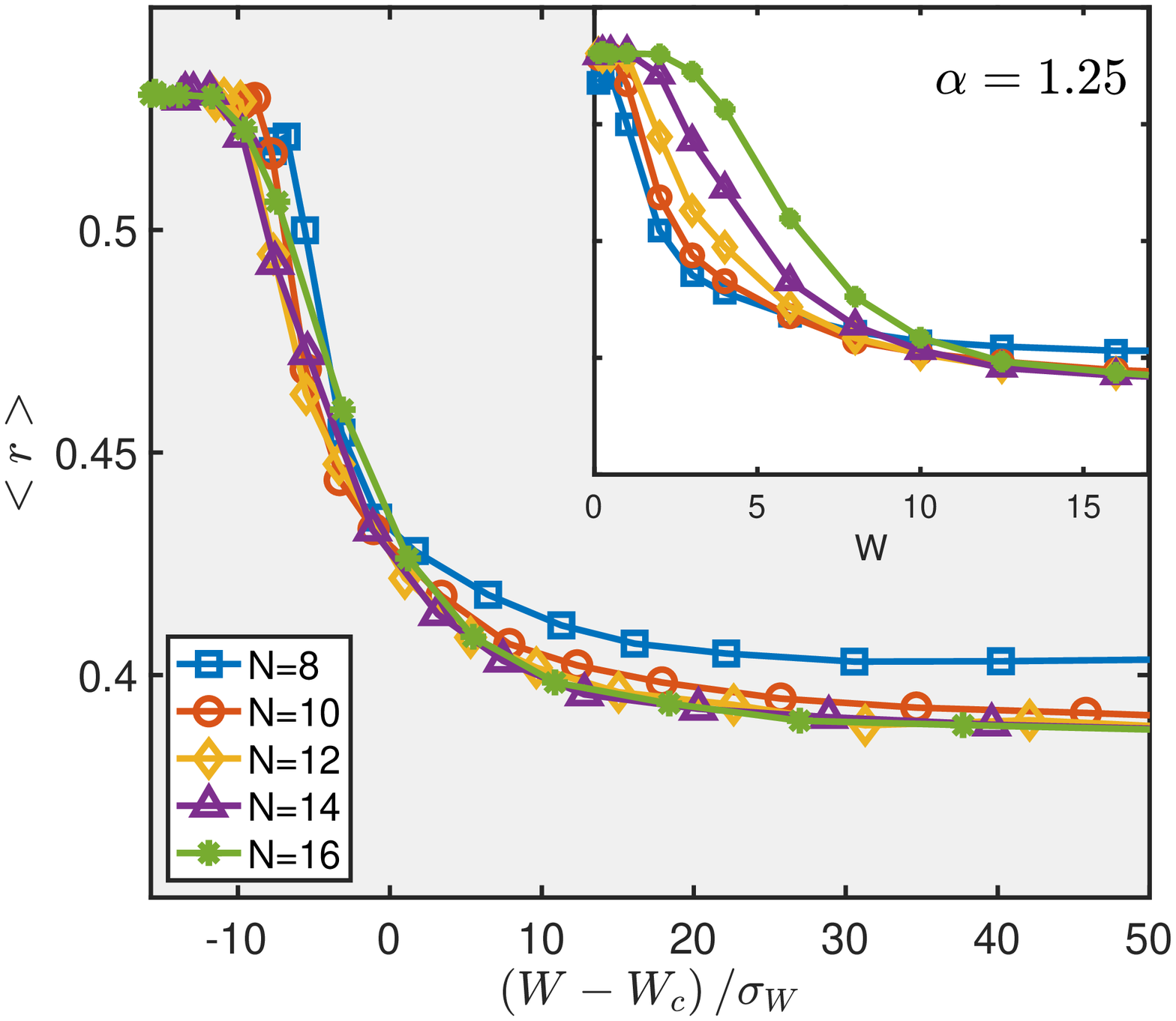}
\caption{\small The level statistics vs disordering rescaled as $(W-W_{c})/\sigma_{W}$ (main plot) with original data (inset) for $8\leq N \leq 16$ and $\alpha=1.25$.}
\label{fig:Stat125}
\end{figure}

\subsubsection{Scaling with power-law exponent}

\begin{figure}[h!]
\centering
\includegraphics[width=0.8\columnwidth]{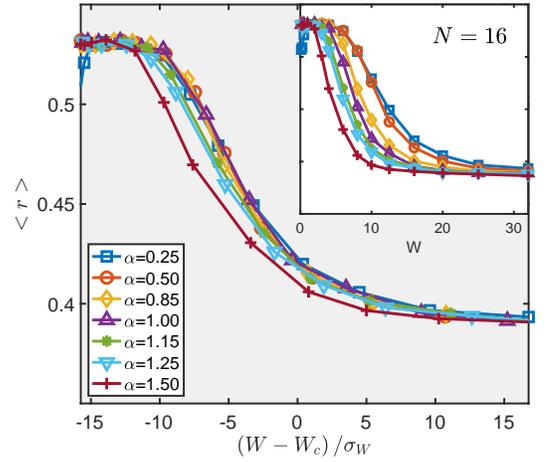}
\caption{\small The level statistics vs disordering rescaled as $(W-W_{c})/\sigma_{W}$ (main plot) with original data (inset) for $0.25 \leq \alpha \leq 3/2$ and  $N=16$.}
\label{fig:StatN16}
\end{figure}

Rescaling for different $\alpha$ and fixed $N$ also gives a nice collapse, as seen on Fig. \ref{fig:StatN16}. It can also be noticed, in contrast with Fig. \ref{fig:HammA16}, that the $\alpha=3/2$ curve went significantly lower than others, but its tail still merged with other curves, which is consistent with the discussion after Eq. \eqref{eq:Case4}; therefore, the results for $\alpha=3/2$ seem to be inconclusive, which confirms its threshold behavior. For $\alpha\leq 1$ there is a nearly perfect match.

%We tried $\alpha>1.5$, and that gave similar results to Ref. \cite{ab15MBLXY}. Constant $W_c(N)$ gives better fitting results for short-range interactions \cite{Luitz2015}.

\section{Conclusion}
\label{sec:Concl}

The proposed model of multispin resonances provides a scaling of the critical disorder corresponding to the MBL transition in the model of spins in random fields coupled by transverse interaction for $0<\alpha<3/2$. It is shown that delocalization takes place due to interacting resonant quartets. We predicted the critical disordering $W_c$ to behave as
\begin{align}
W_c =
\begin{cases}
    \frac{1.37 \alpha}{4/3-\alpha} J_0 N^{4/3-\alpha} \ln N, & \text{$0 < \alpha \leq 1$},\\
    \frac{\alpha}{1-2\alpha/3} J_0 N^{1-2\alpha/3} \ln^{2/3} N, & \text{$1 < \alpha < \frac{3}{2}$},
  \end{cases}  
\end{align} 
and the width of transition as
\begin{align}
\sigma_W \approx \frac{W_c}{N}.
\end{align} 
The scaling of the critical disordering has been predicted considering the localization breakdown by interacting resonant spin quartets, while the quantitative definitions of the critical disordering and the transition width were obtained using the numerical analysis of the system eigenstates for system sizes $8\leq N \leq 16$. 

\begin{figure}[h!]
\centering
\includegraphics[width=0.8\columnwidth]{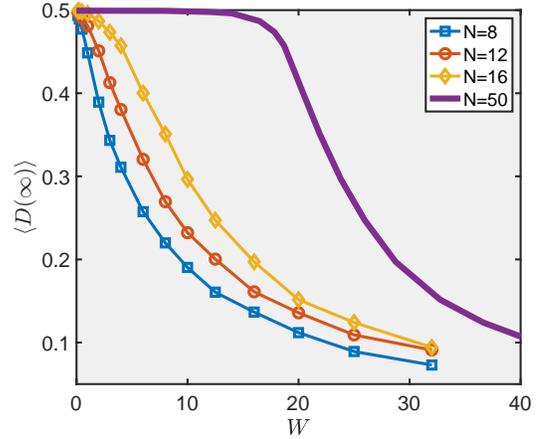}
\caption{\small The theoretically predicted normalized Hamming distance for $N=50$ and $\alpha=1$ comparing to the numerically calculated for $N=8,12,16$.}
\label{fig:Hamm50}
\end{figure}

Based on the obtained scaling, we predict a behavior of the Hamming distance for the system with $N=50$, $\alpha=1$, and $B=4J_0$ described by Eq. \eqref{eq:H} (Fig. \ref{fig:Hamm50}). These results call for comparison with the experiment that can be performed for systems so large \cite{Atoms53Zhang2017,Atoms51Bernien2017}  that they are completely unaffordable for exact diagonalization. 

In the limit $\alpha \rightarrow 0$, the extension of results leads to a full localization, consistent with preliminary studies.

There are no conclusive results for the threshold case $\alpha=3/2$. The logarithmic scaling of the critical disordering, $W_{c}$, gives a satisfactorily data collapse for Hamming distances, while the behavior of level statistics  is more complicated. More sophisticated analytical and numerical studies are needed to describe MBL in this regime.

The analytical expression for the critical disordering can be generalized to higher dimensions, $d>1$, as
\begin{eqnarray}
W_c =c_{\alpha,d}
\begin{cases}
    J_0 N^{4/3-\alpha/d} \ln N, & \text{$0 < \alpha \leq d$},\\
    J_0 N^{1-2\alpha/3d} \ln^{2/3} N, & \text{$d < \alpha < \frac{3d}{2}$}.
  \end{cases}  
  \label{eq:Case4d} 
\end{eqnarray}
The reasoning based on spin quartet resonances can be applied for higher dimensions, $d$, leading to a dimensional constraint $\alpha<3d/2$. The proportionality coefficients, $c_{\alpha,d}$, are not provided here due to the much harder numerical studies of the problem through full diagonalization for $d>1$.

The rescaling of the interaction in a one-dimensional system following Eq. (\ref{eq:Resc}) makes the transition size independent in the thermodynamic limit.  This transition is stable with respect to ergodic spots as justified in Sec. \ref{sec:thermlimresc}. The same behavior can be expected in higher dimensions. %The nature of this transition is different from the delocalization transition for $\alpha > 3d/2$ where ergodic spots are crucially important \cite{Huveneers17BrekDwnLoc} while for $\alpha<3d/2$ the transition seems to be similar to the localization transition in the Bethe lattice \cite{AbouChacra73}. 

\begin{acknowledgments}
This work is partially supported by the National Science Foundation (CHE-1462075) and by Tulane University Bridge Fund.

AB acknowledges the Support of MPIPKS (Dresden, Germany) Visitor Program in 2018 and 2019. 

This work used the Extreme Science and Engineering Discovery Environment (XSEDE), which is supported by National Science Foundation grant number ACI-1548562. Specifically, it used the Bridges system, which is supported by NSF award number ACI-1445606, at the Pittsburgh Supercomputing Center (PSC).

We acknowledge Ivan Khaymovich, Markus Heyl, Giuseppe De Tomasi, David Luitz and Jason Sperry for useful discussions.
\end{acknowledgments}

\bibliography{MBL}

\clearpage
\appendix
\section{Spin pairs}
\label{sec:app-pairs}

The results for localization threshold due to spin pairs were obtained only in the case of $\alpha>1$ \cite{ab15MBLXY}.
Our results for spin quartets in the case of $\alpha<1$ cannot be conclusive without understanding the resonance contribution of spin pairs. We begin with the definition of diagonal interaction at arbitrary distance between pairs. This interaction can be written as $U_{ij}\sigma_{i}^{z}\sigma_{j}^{z}$, with 
\begin{eqnarray}
U_{ij}=\sum_{k}\frac{J_{ij}J_{ik}J_{jk}\phi_{i}\phi_{j}}{(\phi_{i}^2-\phi_{k}^2)(\phi_{j}^2-\phi_{k}^2)}. 
\label{eq:IntDiag}
\end{eqnarray}
The typical interaction depends on the distance between resonant pairs, $R$, and their energy difference, $\phi_{i}-\phi_{j}$. The minimum distance, $R$, and the minimum energy difference, $\varepsilon$,  can be expressed in terms of the total number of resonances, $n_{r}$ as $R \sim N/n_{r}$, and $\varepsilon \sim W/n_r$. In the practically significant case of $n_{r} \sim 1$, one gets $R \sim N$ and $\varepsilon \sim W$, so the interaction is defined by Eq. \eqref{eq:Ulong} as
 \begin{eqnarray}
U_{ij} & \sim J_0^3 W^{-2} N^{-2\alpha}, & ~ \alpha\geq 1;
\nonumber\\
 U_{ij} & \sim J_0^3 W^{-2} N^{1-3\alpha}, & ~ \alpha < 1. 
\label{eq:IntDiagx1alph}
\end{eqnarray}

If $N>n_{r}>1$, one can consider the contribution from either closest distance $R \sim N/n_{r}$, and typical energies $\phi_{i}-\phi_{j} \sim W$, or smallest energy difference $\phi_{i}-\phi_{j} \sim W/n_{r}$, and typical distances $R \sim N$. These contributions can be estimated as 
  \begin{eqnarray}
U_{ij} &\sim \frac{J_0^3n_{r}^{2\alpha}}{W^2N^{2\alpha}}, & ~ 1 \leq \alpha < \frac{3}{2}, ~{\rm small ~ distances},
\nonumber\\
U_{ij} & \sim \frac{J_0^3n_{r}^{2-\alpha}}{W^2N^{2\alpha}}, & ~ 1 \leq \alpha < \frac{3}{2}, ~{\rm small ~ energies},
\nonumber\\
 U_{ij} & \sim \frac{J_0^3 n_{r}^{3\alpha-1}}{W^2N^{3\alpha-1}}, &  ~ \frac{1}{2} <\alpha < 1, ~{\rm small ~ distances},
 \nonumber\\
 U_{ij} & \sim \frac{J_0^3n_{r}}{W^2N^{3\alpha-1}}, & ~ \frac{1}{2}< \alpha < 1, ~{\rm small ~ energies},
 \nonumber\\
  U_{ij} & \sim \frac{J_0^3n_{r}^{\alpha}}{W^2N^{3\alpha-1}}, & ~ 0<\alpha< \frac{1}{2}, ~{\rm small ~ distances},
 \nonumber\\
 U_{ij} & \sim \frac{J_0^3n_{r}}{W^2N^{3\alpha-1}}, & ~ 0 <\alpha< \frac{1}{2}, ~{\rm small ~ energies}.
\label{eq:IntDiagx1alph1}
\end{eqnarray}

The case of $\alpha>2/3$ is determined by small distance regime. For $\alpha>1$, considering pairs of the maximum size $N$, we get $n_{r} = J_0 N^{2-\alpha}/W$; then, setting $U_{ij} \sim V_{ij}=J_0/N^{\alpha}$, we obtain the criterion $W \sim J_0 N^{\frac{\alpha (3-2\alpha)}{2(\alpha+1)}}$, identical to the earlier work \cite{ab15MBLXY}. Similar arguments in the case of $2/3\leq \alpha\leq 1$ yield $W \sim J_0 N^{\frac{5\alpha-3\alpha^2-1}{3\alpha+1}}$.  
The case of $\alpha < 2/3$ is determined by the small energy regime; the calculations for this regime yields $W_{c} \sim J_0 N^{1-\alpha}$, which is always smaller than the contribution of quartets. 

In all regimes the delocalization is determined by quartets, because the diagonal interaction of pairs is too weak to disturb resonances \cite{Burin2017}.

\section{Spin quartets}
\label{sec:app-quartets}

We consider the case of $B=0$, meaning that even for the finite transverse field the threshold disordering satisfies $B \ll W_c$, so the resonance condition for a spin quartet can be written as $\phi_1+\phi_2+\phi_3+\phi_4=0$.
The contribution from spin quartets in the second order gives zero, because only possible flips are independent like
\begin{align}
       \left|\uparrow\uparrow\uparrow\uparrow\right> \rightarrow \left|\downarrow\downarrow\uparrow\uparrow\right> \rightarrow \left|\downarrow\downarrow\downarrow\downarrow\right>,
\end{align}
and they interfere destructively with each other, resulting in zero, as can be seen from the sum over all processes like (here the first spin flips with all other in arbitrary order, and three other processes like that should be added)
\begin{multline}
     V_4^{(2)} = \frac{J_{12}J_{34}}{\phi_1+\phi_2} + \frac{J_{13}J_{24}}{\phi_1+\phi_3} + \frac{J_{14}J_{23}}{\phi_1+\phi_4} +\frac{J_{34}J_{12}}{\phi_3+\phi_4} + \\ + \frac{J_{24}J_{13}}{\phi_2+\phi_4} + \frac{J_{23}J_{14}}{\phi_2+\phi_3}
     = \frac{J_{12}J_{34}}{\phi_1+\phi_2} + \frac{J_{13}J_{24}}{\phi_1+\phi_3} + \\ + \frac{J_{14}J_{23}}{\phi_1+\phi_4} - \frac{J_{34}J_{12}}{\phi_1+\phi_2} - \frac{J_{24}J_{13}}{\phi_1+\phi_3} - \frac{J_{23}J_{14}}{\phi_1+\phi_4} = 0.
\end{multline}
In the third order there is a non-zero contribution from six processes led by first spins (similarly one can consider $18$ more processes led by second, third and fourth spins):
\begin{align}
       \left|\uparrow\uparrow\uparrow\uparrow\right> \rightarrow \left|\downarrow\downarrow\uparrow\uparrow\right> \rightarrow \left|\uparrow\downarrow\downarrow\uparrow\right> \rightarrow \left|\downarrow\downarrow\downarrow\downarrow\right>, \\
      \left|\uparrow\uparrow\uparrow\uparrow\right> \rightarrow \left|\downarrow\uparrow\downarrow\uparrow\right> \rightarrow \left|\uparrow\downarrow\downarrow\uparrow\right> \rightarrow \left|\downarrow\downarrow\downarrow\downarrow\right>, \\
       \left|\uparrow\uparrow\uparrow\uparrow\right> \rightarrow \left|\downarrow\uparrow\uparrow\downarrow\right> \rightarrow \left|\uparrow\downarrow\uparrow\downarrow\right> \rightarrow \left|\downarrow\downarrow\downarrow\downarrow\right>, \\
       \left|\uparrow\uparrow\uparrow\uparrow\right> \rightarrow \left|\downarrow\downarrow\uparrow\uparrow\right> \rightarrow \left|\uparrow\downarrow\uparrow\downarrow\right> \rightarrow \left|\downarrow\downarrow\downarrow\downarrow\right>, \\
       \left|\uparrow\uparrow\uparrow\uparrow\right> \rightarrow \left|\downarrow\uparrow\uparrow\downarrow\right> \rightarrow \left|\uparrow\uparrow\downarrow\downarrow\right> \rightarrow \left|\downarrow\downarrow\downarrow\downarrow\right>, \\
       \left|\uparrow\uparrow\uparrow\uparrow\right> \rightarrow \left|\downarrow\uparrow\downarrow\uparrow\right> \rightarrow \left|\uparrow\uparrow\downarrow\downarrow\right> \rightarrow \left|\downarrow\downarrow\downarrow\downarrow\right>.
    \end{align}
The resulting amplitude, $A_{1,234}$, of contributions from processes led by the first spin can be evaluated as
\begin{multline}
        A_{1,234}=\frac{J_{12}J_{13}J_{14}}{\left(\phi_1+\phi_2\right)\left(\phi_2+\phi_3\right)}+\frac{J_{13}J_{12}J_{14}}{\left(\phi_1+\phi_3\right)\left(\phi_2+\phi_3\right)} +\\ +\frac{J_{14}J_{12}J_{13}}{\left(\phi_1+\phi_4\right)\left(\phi_4+\phi_2\right)}
        +\frac{J_{13}J_{14}J_{12}}{\left(\phi_1+\phi_3\right)\left(\phi_3+\phi_4\right)}+\\ + \frac{J_{12}J_{14}J_{13}}{\left(\phi_1+\phi_2\right)\left(\phi_2+\phi_4\right)}+\frac{J_{14}J_{13}J_{12}}{\left(\phi_1+\phi_4\right)\left(\phi_4+\phi_3\right)} = \\
        =J_{12}J_{13}J_{14}\frac{\left(2\phi_2+\phi_3+\phi_4\right)}{\left(\phi_1+\phi_2\right)\left(\phi_2+\phi_3\right)\left(\phi_2+\phi_4\right)} +\\
        +J_{12}J_{13}J_{14}\frac{\left(2\phi_3+\phi_2+\phi_4\right)}{\left(\phi_1+\phi_3\right)\left(\phi_2+\phi_3\right)\left(\phi_3+\phi_4\right)}
        +\\+J_{12}J_{13}J_{14}\frac{\left(2\phi_4+\phi_2+\phi_3\right)}{\left(\phi_1+\phi_4\right)\left(\phi_2+\phi_4\right)\left(\phi_3+\phi_4\right)} =\\
        =-J_{12}J_{13}J_{14}\frac{4\left(\phi_2+\phi_3+\phi_4\right)}{\left(\phi_2+\phi_3\right)\left(\phi_2+\phi_4\right)\left(\phi_3+\phi_4\right)}.
    \end{multline}

%The amplitude of this process "led" by the first spin is proportional to the product of three coupling constants, $J_{12}J_{13}J_{14}$. The processes with other "leading" spins will be proportional to different products, so they cannot suppress the transition amplitude, except for the case of all equal amplitudes ($\alpha=0$), where the full compensation takes place.
This expression is used in the main text as Eq. \eqref{eq:Aklmn}. The sum of all contributions, $V_{1234}=A_{1,234}+A_{2,341}+A_{3,412}+A_{4,123}$, is not zero due to different products of coupling constants in each contribution, except for the case of all equal amplitudes ($\alpha=0$), where the full compensation takes place.

\end{document}